\documentclass[%
 reprint,
nofootinbib,
nobibnotes,
bibnotes,graphicx,graphics,
 amsmath,amssymb,
 aps,stmaryrd,
]{revtex4-1}
\usepackage{natbib} % bib
\usepackage{soul}
\newcommand{\comment}[1]{}
\usepackage{tikz}
  \newlength\squareheight
\usepackage{wasysym} % pentagon
\usepackage{graphicx}% Include figure files
\usepackage{dcolumn}% Align table columns on decimal point
\usepackage{bm}% bold math
\begin{document}

\preprint{Draft}

\title{Two-pathogen model with competition on clustered networks}

\author{Peter Mann}
\email{pm78@st-andrews.ac.uk}
\author{V. Anne Smith}%
\author{John B.O. Mitchell}
\author{Simon Dobson}
\affiliation{School of Computer Science, University of St Andrews, St Andrews, Fife KY16 9SX, United Kingdom }
\affiliation{EaStCHEM School of Chemistry \& BSRC, University of St Andrews, St Andrews, Fife KY16 9ST, United Kingdom }
\affiliation{School of Biology, University of St Andrews, St Andrews, Fife KY16 9TH, United Kingdom }

\date{\today}% It is always \today, today,
             %  but any date may be explicitly specified

\begin{abstract}
Networks provide a mathematically rich framework to represent social contacts sufficient for the transmission of disease. Social networks are often highly clustered and fail to be locally tree-like. In this paper, we study the effects of clustering on the spread of sequential strains of a pathogen using the generating function formulation under a complete cross-immunity coupling, deriving conditions for the threshold of coexistence of the second strain. We show that clustering reduces the coexistence threshold of the second strain and its outbreak size in Poisson networks, whilst exhibiting the opposite effects on uniform-degree models. We conclude that clustering within a population must increase the ability of the  second wave of an epidemic to spread over a network. We apply our model to the study of multilayer clustered networks and observe the fracturing of the residual graph at two distinct transmissibilities.
\end{abstract}

\pacs{Valid PACS appear here}% PACS, the Physics and Astronomy
                             % Classification Scheme.
%\keywords{Suggested keywords}%Use showkeys class option if keyword
                              %display desired
\maketitle

%\tableofcontents

\section{Introduction}
\label{sec:introduction}

Complex networks can be found across many different areas of biology, medicine, the physical and computer sciences. Each network, empirical or synthetic, has a rich structure that exhibits large-scale emergent properties from local interactions. Amongst their applications, complex networks have proven to be excellent models of social networks. The nodes of the graph represent individuals while the edges that connect them represent points of contact. 

At the time of writing the 2020 \textsc{covid-19} pandemic is still raging and presenting the threat of a ``second wave'' of potentially varying strains of the original SARS-CoV-2 virus.
A significant use of social network dynamics is in the study of epidemic diseases, where infected individuals transmit infection to their social contacts with some probability~\cite{PhysRevE.66.016128,citekey1,PhysRevE.76.036113,PhysRevE.76.010101}.
While the study of single-disease epidemics has a substantial literature, it is important to remember that diseases are organisms evolving under selection pressures.
These different strains of disease can interact with each other in complex ways: the first disease may render an individual immune to a later strain, or make them more susceptible, or indeed be a necessary precursor to later infection.
(All these possibilities -- and more -- can be found in nature.)
Essentially a second disease is introduced into a system that has been equilibriated by the passage of a first disease. 
One way to think about this is that the ``first wave'' of a disease changes the topology of the substrate network over which any ``second wave'' propagates, by changing the population, connectivity, and susceptibility of individuals exposed to later infection.
It is known that disease interactions can introduce non-trivial changes in threshold behaviour of both pathogens.
The case in which the first disease provides complete immunity against the second was studied by Newman~\cite{newman_2005,PhysRevE.84.036106} in the case of purely tree-like networks.
Of particular current interest, however, is the behaviour of multiple strains on clustered networks that more accurately model human contact dynamics, especially in those cases that can lead to co-existence of two pathogens within the network.
An improved understanding of strain dynamics on human contact networks in vital in facilitating the detailed study of countermeasures to limit and control further outbreaks.
The interactions between strains will be critical in determining how prior infection affects future transmission, both directly and through topological changes.

Perhaps the most fundamental network model is the Erd\H{o}s-R\'enyi random graph, a member of the exponential random graph ensemble with a constraint on the number of edges within a given realisation. Random graphs are well studied within the network science community using a variety of mathematical tools. One such theoretical framework, the generating function formulation \cite{PhysRevE.64.026118}, has excellent ability to extract the properties of diseases, such as the number of individuals who become infected,  spreading over such networks. This is achieved by an isomorphism between the spreading pathogen and the bond percolation process. The latter, a model that traces its roots to statistical mechanics, examines the probability that each edge in the network transmits the disease between two neighbours with transmission probability $T_1\in [0,1]$, or fails with probability $1-T_1$. We call edges that transmit the disease \textit{occupied}, while those that do not are said to be \textit{unoccupied}.  Once all edges have been considered, the network may no longer be well connected by the occupied edges. Within the context of the isomorphism, the size of the giant connected component (GCC) among occupied edges represents the fraction of the network that becomes infected by the disease. The expectation value of the GCC experiences a second-order phase transition as a function of $T_1$ at some critical  value, $T_{1,\text{c}}$, known as the epidemic threshold. Prior to the threshold, there is no GCC and only small components are connected.  

Social networks tend to contain a high density of triangles; connections between the neighbours of a node, also known as \textit{transitivity} or clustering. Many mathematical models fail to describe the impact of clustering, which is well known to alter the properties of both bond percolation and the epidemic outbreaks of a single disease. Specifically, it can be shown that clustering reduces the epidemic threshold for the disease to infect a finite fraction of the network as well as reducing the overall outbreak size \cite{PhysRevLett.103.058701} for fixed mean degree. Miller \cite{miller_2009,citekey2} conversely showed that clustering can also increase the threshold when degree-assortativity within the networks is also studied, a result supported by Volz \textit{et alia}~\cite{10.1371/journal.pcbi.1002042}. 

Clustering has been well studied in the context of the generating function formulation for a single strain; it requires a generalisation of the generating function formulation to partition edges into distinct topological sets \cite{miller_2009,PhysRevLett.103.058701}. The random clustered graphs we consider here are built using the generalised configuration model \cite{PhysRevE.82.066118,2020arXiv200606744M,mann2020random}. In this model, a vector of edge-topologies, $\bm \tau$, is defined; the simplest model consists of tree-like edges, denoted by $\bot$ and triangles, denoted by $\Delta$, such that $\bm \tau=\{\bot,\Delta\}$. Each node is assigned a stub-degree, $k_\tau$, for each topology in the topology set, $\tau\in \bm \tau$. For instance, a node involved in 3 tree-like edges and 1 triangle has $k_\bot = 3$ and $k_\Delta=2$ and it should be clear that $\{k_\Delta = 0 \text{ mod } 2\}$. During the network construction, the stubs are connected together to create a random graph whose edge topologies are distributed according to the assigned stub-degree. 

It is not clear, however, precisely how clustering impacts the spread of two cross immune pathogens spreading sequentially over a network. The subject has been studied before using percolation in the context of clique random networks whereby each strain spreads on a particular edge topology \cite{WANG2012121}. In this paper, we study the influence of clustering on the outbreak size of two sequential pathogens spreading with a perfect cross-immune coupling on a random clustered network.  

\section{Sequential strain model with clustering}
\label{sec:model}

In this section, we introduce a two-strain model on clustered networks containing triangles in addition to the tree-like degrees. The second strain is assumed to temporally separated from the first such as seasonal influenza outbreaks or a rare mutation in an equilibrated bacterial population.

\subsection{Strain-1}
\label{subsec:S1}

The generating function formulation \cite{PhysRevE.64.026118,PhysRevE.66.016128} rests upon the degree distribution, $p(k)$, the probability of choosing a node at random from the network of degree $k$. When the network contains triangles, we introduce the joint degree distribution, $p(k_\bot, k_\Delta)$, the probability of choosing a node at random from the network with $k_\bot$ tree-like edges and $k_\Delta/2$ triangles. We can recover $p(k)$ from the joint degree sequence as 
\begin{equation}
    p(k) = \sum^\infty_{k_\bot=0}\sum^\infty_{k_\Delta=0}p(k_\bot,k_\Delta)\delta_{k,k_\bot +k_\Delta}
\end{equation}
The joint probability distribution is generated by 
\begin{equation}
    G_0(z_\bot,z_\Delta) = \sum^\infty_{k_\bot=0}\sum^\infty_{k_\Delta=0} p(k_\bot,k_\Delta){z_\bot}^{k_\bot}{z_\Delta}^{k_\Delta/2}
\end{equation}
The probability of reaching a node of joint degree $(k_\bot,k_\Delta)$ by following a random tree-like edge back to a node is generated by 
\begin{equation}
    G_{1,\bot}(z_\bot,z_\Delta) = \frac{1}{\langle k_\bot \rangle}\frac{\partial G_0}{\partial z_\bot}
\end{equation}
Similarly, the degree of the node reached by following a random triangle edge to a node is 
\begin{equation}
    G_{1,\Delta}(z_\bot,z_\Delta) = \frac{1}{\langle k_\Delta \rangle}\frac{\partial G_0}{\partial z_\Delta}
\end{equation}
In each case, $\langle k_\tau\rangle$ is the average $\tau$-degree of a node which is given by $\partial_{z_\tau} G_0(1,1)$. 

The clustering coefficient $\mathcal C$ is a metric that indicates the level of clustering in the network \cite{PhysRevLett.103.058701, PhysRevE.74.056114}. It is given by the following quotient
\begin{equation}
    \mathcal C = \frac{3N_\Delta}{N_3}
\end{equation}
where $N_\Delta$ is the number of triangles and $N_3$ is the number of connected triples. In terms of the above generating functions and network size $N$, we have 
\begin{align}
    3N_\Delta =\ & N \left(\frac{\partial G_0}{\partial z_\Delta}\right)\\
    N_3=\ &\frac 12 N\sum_{k=0}^\infty \binom{k}{2}p_k
\end{align}

The probability that a node does \textit{not} become infected through its involvement in a tree-like edge (triangle) is $g_\bot$ $(g_\Delta)$. Each $g_\tau$ is a function of $u_\tau$, the probability that a neighbour is uninfected in a $\tau$-site. These expressions are well-known for both tree-like and triangle edge topologies.
\begin{figure}
\begin{center}
\includegraphics[width=0.35\textwidth]{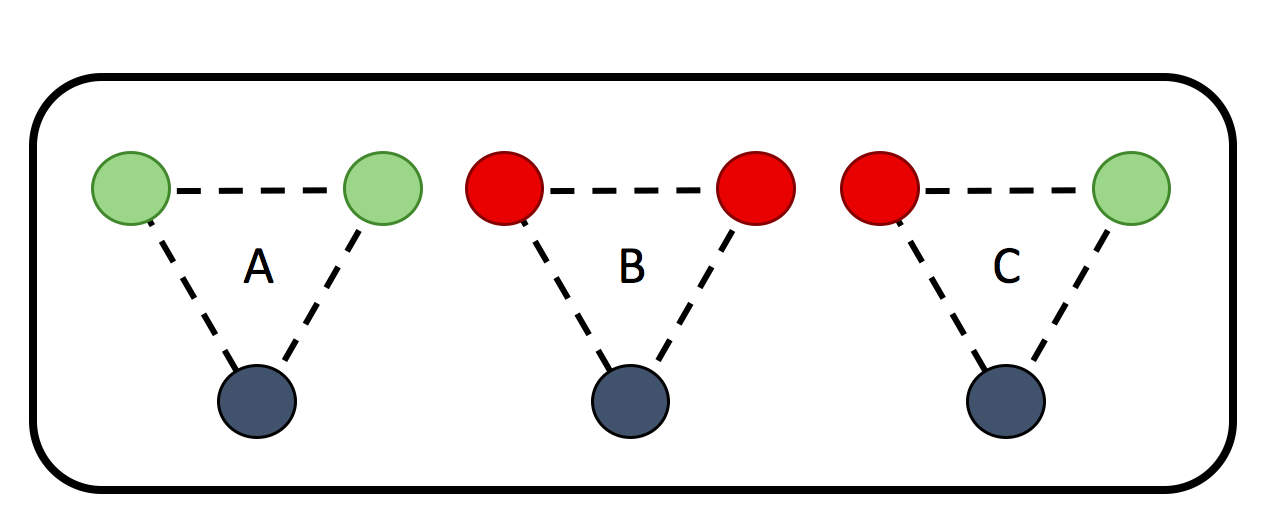}  
\caption[3triangles]{ 
 The three triangles that a focal node may be connected to. (A) The focal node has two uninfected neighbours (green), neither of which are capable of transmitting infection. (B) Both nodes are infected (red), but each direct edge fails to infect the focal node. (C) Only one neighbour is infected; however, it can infect the focal node by first infecting the susceptible neighbour and then a further transmission to the focal node.
} \label{fig:3triangles}
\end{center}
\end{figure}
We construct $g_\bot(u_\bot;T_1)$ by summing the independent probabilities that a given tree-like edge fails to infect the focal node; this is either because the neighbouring node was uninfected by the disease with probability $u_\bot$, or that it was infected but failed to transmit the disease to the focal node with probability $(1-u_\bot)(1-T_1)$. Together we have 
\begin{equation}
    g_\bot(u_\bot;T_1) = u_\bot+(1-u_\bot)(1-T_1)\label{eq:gbot}
\end{equation}

The $g_\Delta(u_\Delta;T_1)$ expression is slightly more complex to consider due to the inter-neighbour connecting edge. For a node that has $\eta_\Delta$ triangles (and therefore has triangle degree $k_\Delta=2\eta_\Delta$), there are three ways to consider the failure to infect the focal node as in Fig \ref{fig:3triangles}. 

Firstly, both neighbours can themselves be uninfected with probability $u_\Delta ^2$. Similarly, both neighbours could have been infected but both failed to transmit their infection to the focal node directly with probability $[(1-u_\Delta)(1-T_1)]^2$; in this case the inter-neighbour edge has no consequence on the final state of the focal node. However, in the case that one neighbour is infected, fails to transmit directly to the focal node and the other node is initially uninfected (the probability of which is $u_\Delta(1-u_\Delta)(1-T_1)$), then the inter-neighbour edge can be an avenue of infection back to the focal node. The probability that this fails to occur is $1-T_1^2$. Allowing there to be $\eta_\Delta$ triangles around the focal node we have
\begin{widetext}
\begin{equation}
   g_\Delta(u_\Delta;T)= \binom{\eta_\Delta}{l}[u_\Delta^2]^l\binom{ \eta_\Delta-l}{m} [((1-u_\Delta)(1-T_1))^2]^m[2u_\Delta(1-u_\Delta)(1-T_1)(1-T_1^2)]^{\eta_\Delta-l-m}\label{eq:gtriangle}
\end{equation}
\end{widetext}
The multiplication by two in the final term is due to the symmetry of the triangle. Each square bracket contains the probability that the focal node remains uninfected in the particular triangle it is considered to be a part of. 

To solve for the expected fraction of the network that contracts strain-1, $S_1$, we use fixed-point iteration to find each $u_\tau$ value as the solution to a self-consistent functional equation in $u_\tau$
\begin{equation}
    u_\tau = G_{1,\tau}(g_\bot,g_\Delta)\label{eq:G1_1}
\end{equation}
each equation converging on a solution in the unit interval. With these values, $S_1$ can be found by solving 
\begin{equation}
    S_1[u_\bot,u_\Delta;T_1] = 1 - G_0(g_\bot,g_\Delta)\label{eq:disease1}
\end{equation}
where the square brackets indicate the functional dependency of the GCC on $u_\tau$ and the disease transmission parameter, $T_1$.

\subsection{Strain-2}
\label{subsec:S2}

Once the first strain has passed through the network, a fraction, $S_1$, of the nodes will have contracted it and consequently a fraction, $1-S_1$, remained uninfected. In the case that nodes infected by strain 1 have perfect cross immunity against further strains, then only those nodes in the fraction $1-S_1$, termed the \textit{residual graph} (RG), can become infected by the second strain. The threshold criterion for the emergence of the second strain on unclustered random graphs has been solved previously by Newman. We now proceed to understand the role of clustering on the second strain. 

Setting the transmissibility of the second strain to $T_2$, the probability that the second strain fails to infect a node chosen at random is comprised of the probabilities that both the tree-like edges and the triangle edges each fail to transmit the strain. In analogy to the first disease, we define the probability $h_\bot$ to be the probability that a tree-like edge remains unoccupied following both strains and introduce $v_\bot$ is the probability that a neighbouring node at the end of a tree-like contact does not have disease 2. The probability that a node with $k$ tree-like contacts has precisely $l\leq k$ susceptible neighbours following disease 1 of which $m\leq l$ also failed to contract disease 2 is given by 
\begin{widetext}
\begin{align}
   h_{\bot}(u_\bot,v_\bot;T_1,T_2) =\ & \binom{k}{l} \binom{l}{m}[u_\bot v_\bot]^m[u_\bot(1-v_\bot)(1-T_2)]^{l-m}[(1-u_\bot)(1-T_1)]^{k-l}
\end{align}
\end{widetext}
Similarly, the probability, $h_\Delta$, that a focal node involved in a triangle fails to become infected is given by the probability that each avenue of infection fails, as considered for the first disease in Eq \ref{eq:gtriangle}. Defining $v_\Delta$ to be the probability that a node involved in a triangle, that is also in the RG of the first strain, remains uninfected during the second epidemic, we now examine each bracket in Eq \ref{eq:gtriangle}.

In the first case, both nodes are uninfected with strain-1 with probability $u_\Delta^2$. To remain uninfected with strain-2, these nodes must fail to transmit to the focal node. This can occur in three distinct ways: either both neighbours fail to contract strain-2, $v_\Delta^2$, or they both have disease-2 but fail to transmit, $((1-v_\Delta)(1-T_2))^2$, or finally, one remains uninfected with strain-2 and the other fails directly to infect with probability $2v_\Delta(1-v_\Delta)(1-T_2)$.

Next, in the case when the RG contains both an infected and an uninfected node, there are only two ways that the focal node can remain uninfected by strain-2. These are the probability that the neighbour remains uninfected, $v_\Delta$, or is infected but fails to transmit, $(1-v_\Delta)(1-T_2)$. Together, these terms can be written as
\begin{widetext}
\begin{align}
    h_\Delta(u_\Delta,v_\Delta;T_1,T_2) =\ & \binom{\eta}{l}[u_\Delta^2]^l \binom{l}{j}[v_\Delta^2]^j\binom{l-j}{i}[2v_\Delta (1-v_\Delta)(1-T_2)(1-T_2^2)]^i[((1-v_\Delta)(1-T_2))^2]^{l-j-i}\nonumber\\
    \ &\binom{\eta-l}{m}[2u_\Delta (1-u_\Delta)(1-T_1)(1-T^2)]^m\biggl(\genfrac{}{}{0pt}{}{m}{f}\biggr)[v_\Delta ]^f[(1-v_\Delta)(1-T_2)]^{m-f}\nonumber\\
    \ &[((1-u)(1-T_1))^2]^{\eta-l-m}
\end{align}
Upon application of the binomial theorem this expression becomes 
\begin{align}
    h_\Delta(u_\Delta,v_\Delta;T_1,T_2) =\ &[u_\Delta^2[v_\Delta^2+2v_\Delta(1-v_\Delta)(1-T_2)(1-T_2^2)+[(1-v_\Delta)(1-T_2)]^2]\nonumber\\
    \ & +[2u_\Delta(1-u_\Delta)(1-T_1)(1-T_1^2)[v_\Delta+(1-v_\Delta)(1-T_2)]]+[((1-u_\Delta)(1-T_1))^2]
\end{align}
\end{widetext}
Despite the length of this equation, the interpretation is simple, we spread strain-2 according to the triangle formula of Eq \ref{eq:gtriangle} in the case that the residual motif is a triangle (motif (A) in Fig \ref{fig:3triangles}), we spread according to the tree-like expression when the residual triangle has only one neighbour in the RG (motif C) and finally, we do not spread strain-2 in the case that the motif is completely part of the GCC of strain-1 (motif B) . 
\begin{figure}
\begin{center}
\includegraphics[width=0.48\textwidth]{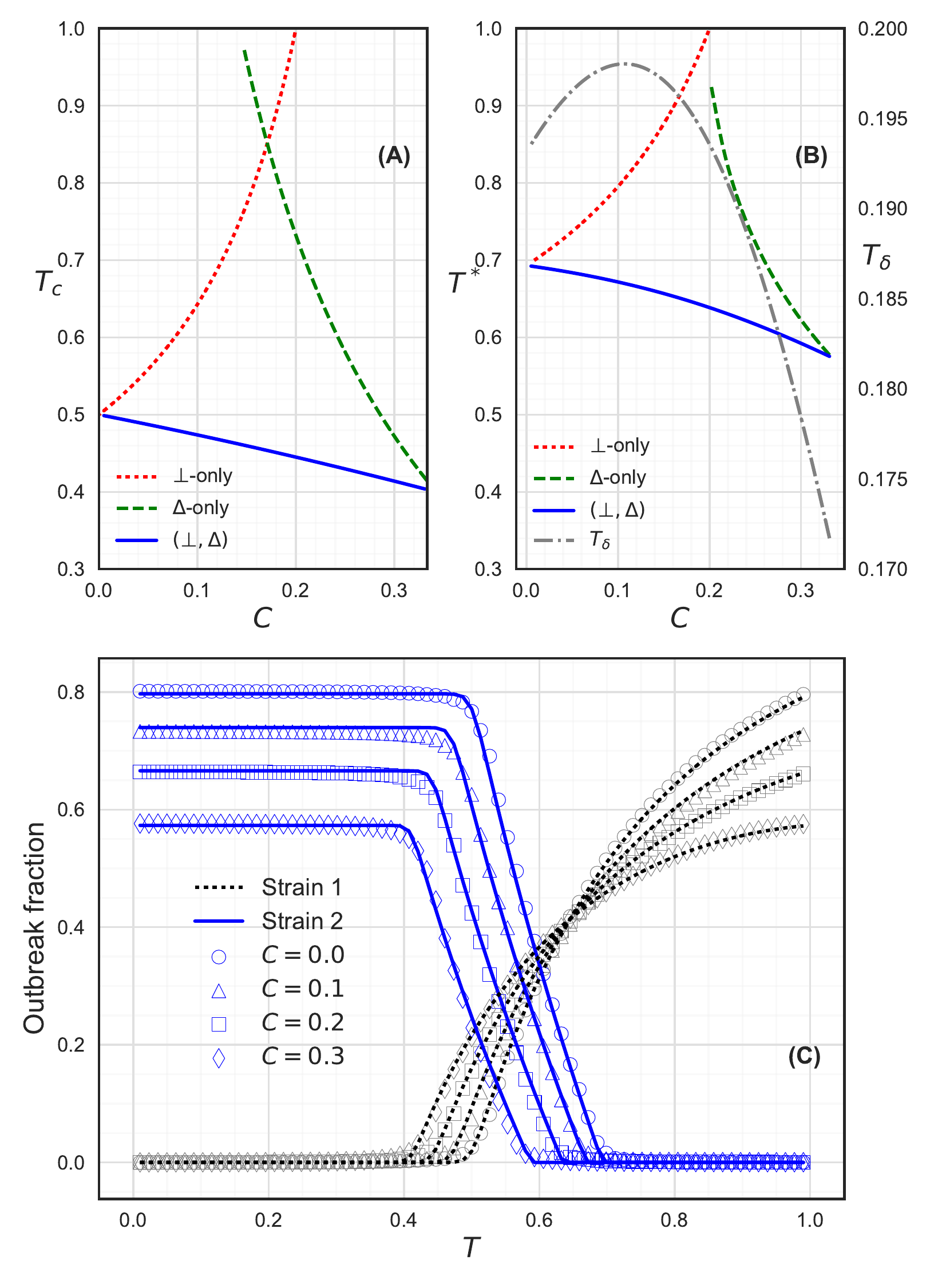}
\caption[together]{ 
 The percolation properties of the 2-strain model over clustered Poisson networks with clustering coefficient, $\mathcal C$, and fixed average degree $\mu+2\nu=2$ of tree-like and triangles, respectively. (A) The epidemic threshold of strain-1 (solid) as a function of $\mathcal  C$.  The critical thresholds for a GCC to exist solely among tree-like edges (small dash) or triangle edges (long dash) from Eq \ref{eq:strain1R0} are plotted in (A). When $C=0$ we have $\nu=0$ indicating the threshold is $T_{1,c}=1/2$, while at $C=1/3$ we have $\mu=0$ and hence find the critical threshold as the root of $T_1^2+2T_1-1=0$ and hence $T_{1,c}\approx 0.41$. Similar analysis in plot (B) shows the coexistence threshold, $T_1^*$, as a function of increasing clustering coefficient from Eq \ref{eq:strain2R0}. Also plotted in (B) is the difference $T_\delta=T_{1,c}-T_1^*$ between the epidemic and coexistence thresholds. After a short increase, $T_\delta$ sharply falls with increased $\mathcal C$, reducing the permissible transmissibilities of strain 1 that allow the coexistence with strain-2. (C) The expected epidemic size of each strain. Scatter points indicate experimental results of bond percolation on a network of size $N=40000$ with $70$ repetitions.  Solid lines represent the theoretical predictions of Eqs \ref{eq:disease1} and \ref{eq:disease2} for each strain. 
} \label{fig:together}
\end{center}
\end{figure}

We can generate $v_\tau$ by writing self-consistent expressions, this time however, dividing by the prior probability that the neighbour does indeed belong to the RG, which is simply $u_\tau$. 
\begin{equation}
    v_\tau = G_{1,\tau}(h_\bot,h_\Delta)/u_\tau\label{eq:G1_2}
\end{equation}
The expectation value for the probability that a randomly chosen node fails to be infected by either strain is
\begin{equation}
     A = \frac{G_0(h_\bot,h_\Delta)}{1-S_1}
\end{equation}
where we have divided by the prior probability of belonging to the RG of disease 1. The fraction of the RG that belongs to the outbreak of the second strain, the giant residual connected component (GRCC), is then given by \begin{equation}
    S_2[u_\tau,v_\tau;T_1,T_2] = (1- A)(1-S_1)\label{eq:disease2}
\end{equation}
The complete prescription is as follows: we use Eq \ref{eq:G1_1} to compute $u_\tau\ \forall \tau\in \bm\tau$, we can then use Eq \ref{eq:disease1} to compute the epidemic outbreak size of the first strain. With these ingredients we calculate $v_\tau\ \forall \tau\in \bm\tau$ using Eq \ref{eq:G1_2} before finalising the calculation of the second outbreak fraction with Eq \ref{eq:disease2}. 

A numerical example of the both strains can be seen in plot (C) of Fig \ref{fig:together} for varying clustering coefficients. The networks for the model are created according to the configuration model \cite{miller_2009,PhysRevLett.103.058701} where the stub-degrees of both tree-like $(k_\bot)$ and triangle $(t=k_\Delta/2)$ topologies of each node are Poisson distributed. The joint degree-distribution is given by \cite{PhysRevLett.103.058701}
\begin{equation}
    p(s,t) = e^{-\mu}\frac{\mu^{k_\bot}}{k_\bot!} e^{-\nu}\frac{\nu^t}{t!} 
\end{equation}
where $\mu$ is the average tree-like degree and $\nu$ is the average number of triangles. The clustering of each network is varied such that the mean degree is fixed at 2. From this we find the means of each Poisson degree sequence as $\mu+2\nu=2$.
As the clustering coefficient increases the epidemic threshold of the first strain decreases from $T_{1,c}=0.5$ to $T_{1,c}
\approx 0.41$. The overall epidemic size at $T_1=1$ is \textit{reduced} as a function of increasing clustering coefficient. Therefore, in this experiment, clustering is seen to have a dual effect on the outbreak of strain-1 depending on $T_1$; clustered networks can expect an epidemic at lower $T_1$, but also expect fewer people to become infected. Setting $T_2=1$, the total outbreak size of the second strain decreases as a function of increased clustering. 

\begin{figure}
\begin{center}
\includegraphics[width=0.5\textwidth]{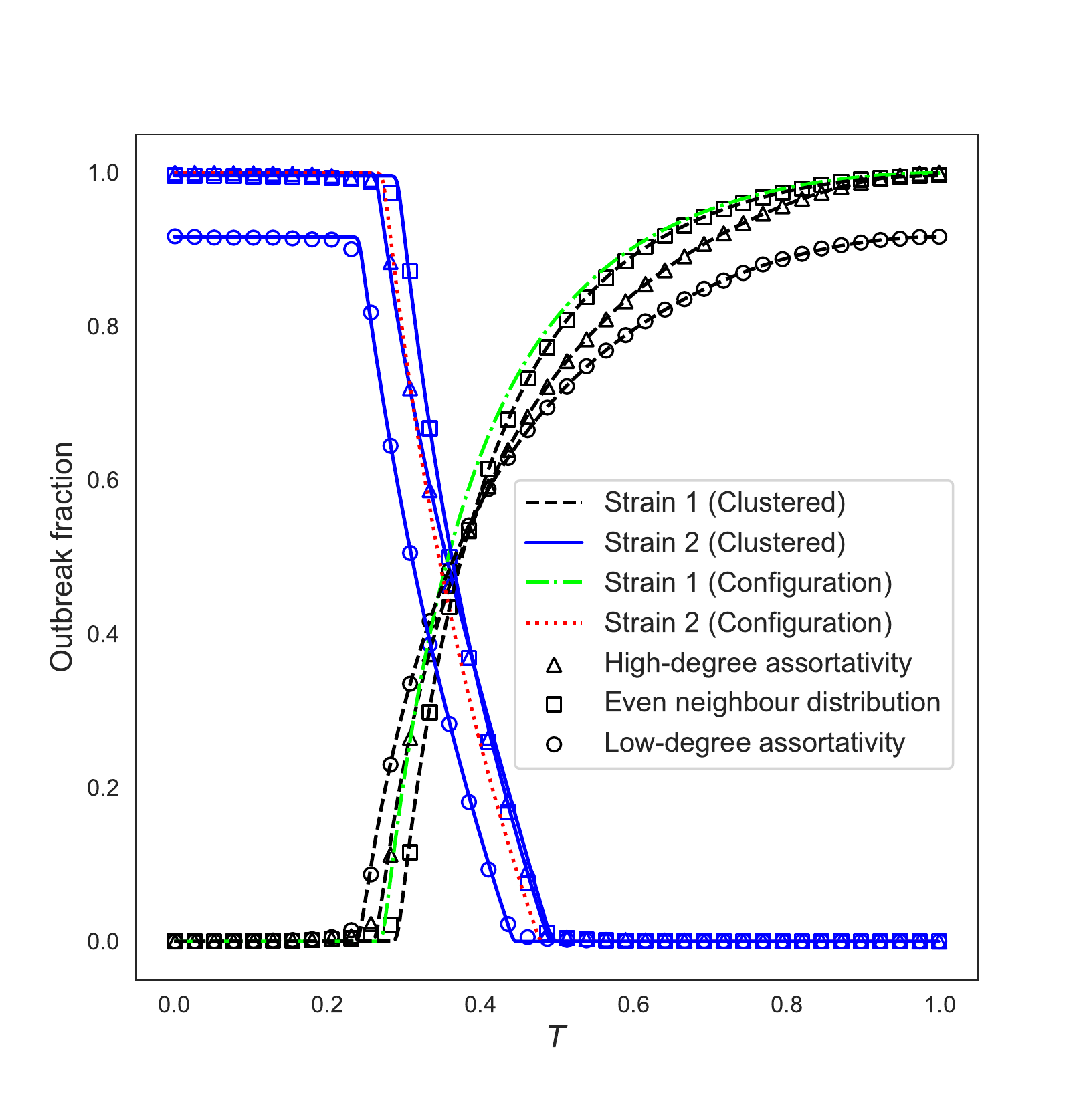}
\caption[miller]{ 
  The residual network of the three networks described in section \ref{subsec:S2}. In this model, clustering can be shown to \textit{increase} both the size of the GRCC and also the coexistence threshold relative to the configuration model. We also observe dichotomous results depending on the nature of the degree assortativity among the clustered edges. When clustering is confined to low-degree nodes, the results of the Poisson experiment are reproduced.
} \label{fig:miller}
\end{center}
\end{figure}

In a second experiment we fix the degrees of each node according to the uniform-degree model, defined in \cite{PhysRevE.80.020901}, enabling the effects of degree-assortativity to be understood. Bond percolation is run on three networks whose nodes have either degrees $2,4$ and $6$, but their clustering is distributed differently. The first has a joint degree distribution of $p(2,0) = 1/3$, $p(2,1)=1/3$ and $p(0,3)=1/3$, increasing the clustering of the high-degree sites. The second network has an even neighbour distribution with $p(2,0)=1/6$, $p(0,1)=1/6$, $p(2,1)=1/3$, $p(4,1)=1/6$ and $p(0,3)=1/6$. Finally, the third network has clustering predominantly among the low-degree sites with $p(6,0)=1/3$, $p(2,1)=1/3$ and $p(0,1)=1/3$. The percolation properties of these networks are presented in Fig \ref{fig:miller}, along with the prediction from the configuration model. In contrast to the random Poisson networks, clustering is shown to increase both the GRCC and  the coexistence threshold relative to the configuration model. Assortativity among low-degree clustered nodes leads to the emergent properties observed by the random Poisson networks.

\subsection{$\mathcal R_0$}
\label{subsec:R0}

The $\mathcal R_0$ value, also known as the case reproduction number of a disease, is a quantity used in epidemiology to represent the number of infections that the average infected node in the network will cause. When the disease has a low transmissibility $T_1\leq T_{1,c}$, we do not expect that an epidemic will occur throughout the entire network, in other words, the infections fizzle out over time. In these cases the $\mathcal R_0$ value is less than unity. $\mathcal R_0=1$ marks the threshold for which the epidemic infects a macroscopic fraction of the population and at this value the transmissibility experiences a critical point, $T_1=T_{1,c}$. Under the bond percolation isomorphism, a GCC of occupied edges forms in the network at and after this bond occupancy probability. The critical transmissibility of the first strain can be found by applying the Molloy-Reed criterion to the configuration model \cite{miller_2009}
\begin{align}
    &\left(\frac{dg_\bot}{du_\bot}\frac{\langle k_\bot^2-k_\bot\rangle}{\langle k_\bot\rangle}-\mathcal R_0\right)\left(\frac{dg_\Delta}{du_\Delta}\frac{\langle k_\Delta^2-k_\Delta\rangle}{\langle k_\Delta\rangle}-\mathcal R_0\right)\nonumber\\
    &=\frac{dg_\bot}{du_\bot}\frac{dg_\Delta}{du_\Delta}\frac{\langle k_\bot k_\Delta\rangle^2}{\langle k_\bot\rangle\langle k_\Delta\rangle}\label{eq:strain1R0}
\end{align}
where each derivative is evaluated at the point $u_\tau=1$. Each bracket on the left hand side can be used to investigate if a GCC occurs among the edges of a given topology; or, the entire expression can be used to determine of the entire network is connected, irrespective of the edge-type, see plot (A) in Fig \ref{fig:together}. It is clear from this plot that clustering increases the interval $T_1\in[T_{1,\text{c}},1]$ by the reduced epidemic threshold, allowing a finite-sized epidemic at lower transmissibilities. 
 
Newman \cite{newman_2005} found that the RG also experiences a phase transition due to the availability of nodes that are not within the GCC as a function of $T_1$. In the case of clustered networks, we find the condition to be given by 
\begin{align}
    &\left(\frac{\partial h_\bot}{\partial  v_\bot}\frac{\langle k_\bot^2-k_\bot\rangle}{\langle k_\bot\rangle}-\mathcal R_0\right)\left(\frac{\partial h_\Delta}{\partial v_\Delta}\frac{\langle k_\Delta^2-k_\Delta\rangle}{\langle k_\Delta\rangle}-\mathcal R_0\right)\nonumber\\
    &=\frac{\partial h_\bot}{\partial v_\bot}\frac{\partial h_\Delta}{\partial v_\Delta}\frac{\langle k_\bot k_\Delta\rangle^2}{\langle k_\bot\rangle\langle k_\Delta\rangle}\label{eq:strain2R0}
\end{align}
The derivatives are evaluated at the point $v_\tau=1$; however we must find the point $(T_1^*,u_\tau^*)$ that satisfies this. As with the first strain, the outbreak size of the second pathogen among only the tree-like or the triangle edges can be found by examining each bracket on the left hand side of Eq \ref{eq:strain2R0}. The emergence of a GCC among the entire RG is found using the entire expression, according to (B) in Fig \ref{fig:together}.

The coexistence threshold, $T_1^*$, for the emergence of a GCC among the tree-like edges of the RG was derived previously by Newman \cite{newman_2005}. Setting $T_2=1$, we find 
\begin{equation}
    \partial _{v_\bot}h_\bot \big|_{v_\bot=1} = u_\bot^*
\end{equation}
and hence the coexistence threshold among tree-like components is
\begin{equation}
    T_1^* = \frac{u_\bot^*-1}{G_{1,\bot}(u_\bot^*)-1}
\end{equation}
The coexistence threshold for the emergence of a GCC among the triangles is slightly harder to solve. Again, with $T_2=1$, we find 
\begin{equation}
     \partial _{v_\Delta}h_\Delta \big|_{v_\Delta=1}  = 2u_\Delta^2+2u_\Delta(1-u_\Delta)(1-T_1)(1-T_1^2)
\end{equation}
For brevity, we use the notation $\kappa = \langle k^2_\Delta-k_\Delta\rangle/\langle k_\Delta\rangle$ and hence we arrive at an equation just in $T_1$
\begin{align}
   & (T_1^*)^3-(T_1^*)^2-T_1^* \nonumber\\
    &+ \frac{2\kappa G_{1,\Delta}(1,g_\Delta^*)-1}{2\kappa G_{1,\Delta}(1,g_\Delta^*)(1-G_{1,\Delta}(1,g_\Delta^*))}=0
\end{align}
where we have used Eq \ref{eq:G1_1} to solve for $u_\Delta$ given $T_1$ in the absence of tree-like edges. 

From plot (B) in Fig \ref{fig:together}, it is clear that the interval $[0, T^*_1]$, which defines the transmissibility range within which strain-2 can exist on the network, is reduced as $T_1^*$ decreases as a function of increasing $\mathcal C$. Comparison of plots (A) and (B) indicates that while both $T_{1,\text{c}}$ and $T_1^*$ fall with $\mathcal C$, the interval $[T_{1,\text{c}},T_1^*]$, which defines the coexistence of each strain on the network, also is reduced, since, $T_1^*$ falls faster than $T_{1,\text{c}}$. This indicates that clustering reduces the total fraction of the population affected at any given $T_1$; decreasing the range of values of $T_1$ at which strain-2 can coexist with strain-1 present; and finally, decreasing the largest value of $T_1$ at which strain-2 is found in the network, squeezing it to a smaller region of the model's phase space.  

\subsection{$C_{\text{RG}}$}
\label{subsec:clusteringcoeff}

The clustering coefficient,  $\mathcal C_{\text{RG}}$, of the RG can be determined and used to investigate how the substrate network fractures following bond percolation. To derive $\mathcal C_{\text{RG}}$ we examine the case when $T_2=1$, and define 
\begin{align}
    j_\Delta(\bm x) =\ &[u_\Delta  x_1]^2 +[(1-u_\Delta)(1-T_1)x_2]^2\nonumber\\
    &+[2u_\Delta(1-u_\Delta)(1-T_1)(1-T_1^2) x_3]
\end{align}
and
\begin{align}
    j_\bot(\bm y) = u_\bot y_1+(1-u_\bot)(1-T_1)y_2
\end{align}
Only $x_1$ retains triangle connectivity in the RG and hence 
\begin{equation}
    3N_{\Delta,\text{RG}} = \frac{\partial }{\partial x_1}\frac{G_0(j_\bot(\bm y),j_\Delta(\bm x))}{G_0(g_\bot,g_\Delta)}\bigg|_{\bm x =\bm y=\bm  1}
\end{equation}
where $G_0(g_\bot,g_\Delta)$ is the prior probability that the node belongs to the RG. The number of connected triples is given by 
\begin{equation}
    N_{3,\text{RG}} = \frac 12 \frac{\partial^2 }{\partial x^2}\frac{G_0(j_\bot(\bm y),j_\Delta(\bm x ))}{G_0(g_\bot,g_\Delta)}\bigg|_{\bm x =\bm y= \bm 1}
\end{equation}
where $z_i=x$ $\forall z_i \in \bm x\cup\bm y$. It now follows that
\begin{equation}
\mathcal C_{\text{RG}}=\frac{3N_{\Delta,\text{RG}} }{N_{3,\text{RG}}}
\end{equation}

\section{Epidemics on Multilayer networks}
\label{sec:multilayer}
We will now apply the 2-strain model to clustered multilayer networks \cite{2020arXiv200606744M}. For simplicity, we consider a 2-layer system comprised of tree-like edges in the first (orange) layer and triangle edges in the second (green) layer. The two layers are connected via interlayer tree-like edges. 

The model is a tautological extension of the model presented in section \ref{sec:model}; strain-2 spreading over the RG created by the GCC of the bilayer networked system. Representing interlayer tree-like edges that an orange (green) node has as $\bot_{\text{og}}$ ($\bot_{\text{go}}$), the vector of permissible topologies is given by $\bm\tau_{\text{o}}=\{\bot_{\text{o}}, \bot_{\text{og}}\}$ for the orange layer and $\bm\tau_{\text{g}}=\{\Delta_{\text{g}}, \bot_{\text{go}}\}$ for the green layer, respectively. Following \cite{sbs3887,2020arXiv200606744M}, each layer has its own $G_{0,\lambda}(\bm z)$ equation, and each element of the topology vectors has its own $G_{1,\lambda,\tau}(\bm z)$ equation also, where $\lambda\in\{\text{o},\text{g}\}$ is a layer index.
\begin{figure}
\begin{center}
\includegraphics[width=0.5\textwidth]{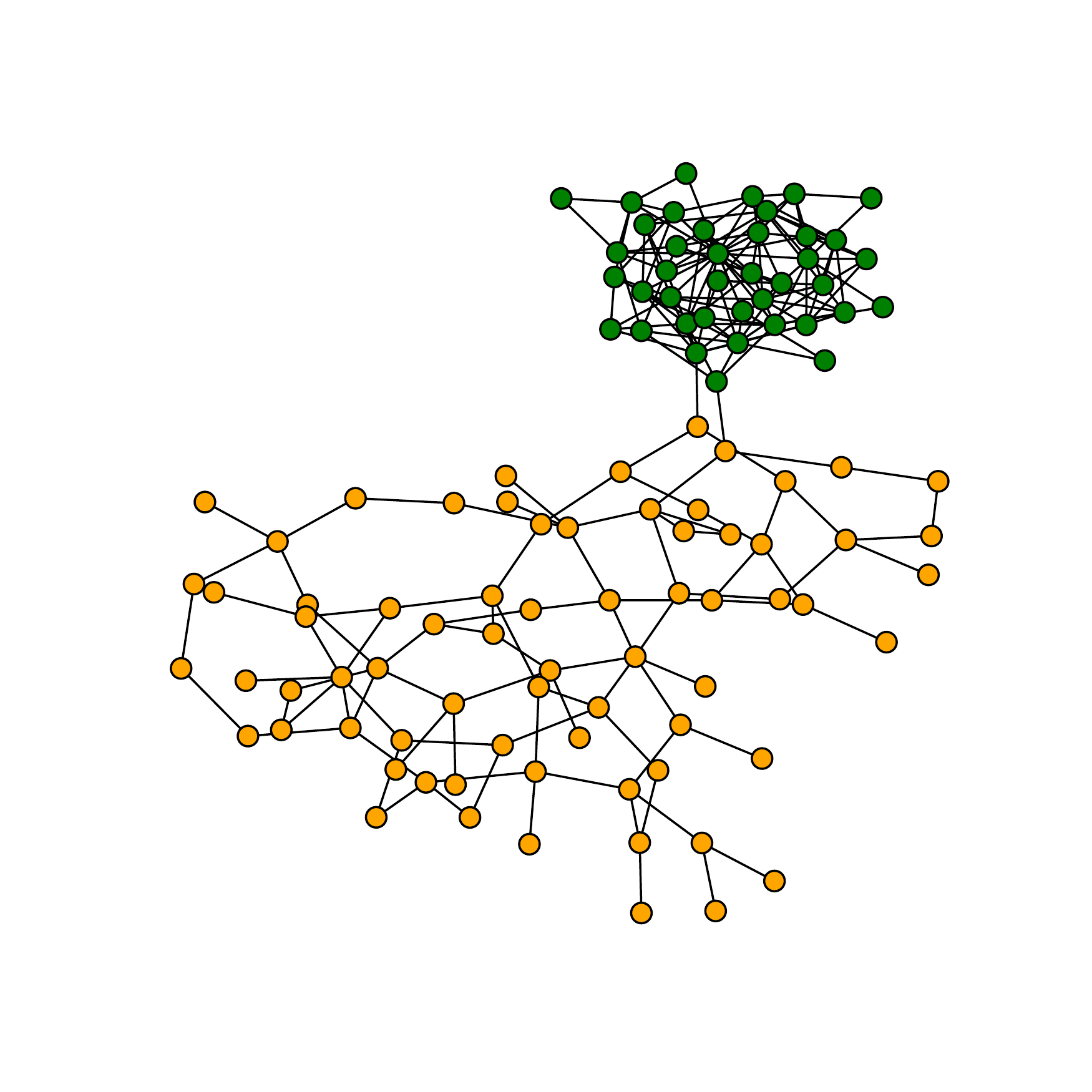}
\caption[multilayernetwork]{ 
An example of the multilayer network used to in the numerical example. The green layer consists solely of triangles while the orange layer is tree-like. Each layer is connected via a few tree-like edges to allow the GCC to span the network. 
} \label{fig:multilayernetwork}
\end{center}
\end{figure}
\begin{figure}
\begin{center}
\includegraphics[width=0.5\textwidth]{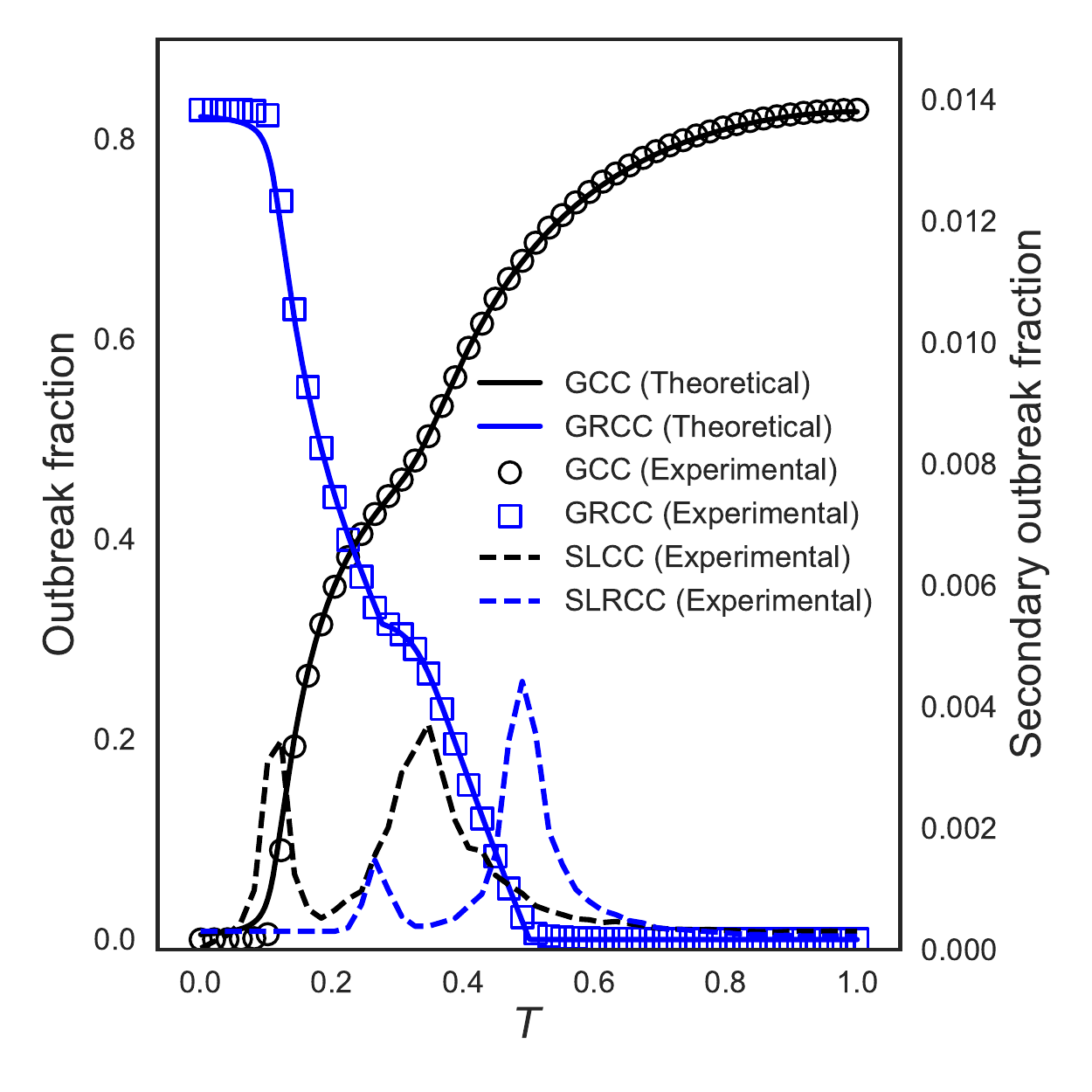}   
\caption[multilayer]{ 
 The expected epidemic size of each strain on a Poisson distributed clustered multilayer network with 2-layers. In this experiment, the orange layer has a clustering coefficient of $C=0$ while the green layer is set to $C=1/3$. Interlayer tree-like edges have been added to allow the GCC to span the entire network. Scatter points indicate experimental results of bond percolation on a network of size $N=20000$ with 25 repeats. Solid lines represent the theoretical predictions of Eqs. Also plotted is the SLCC and the SLRCC, peaks in which indicate a phase transition. From this plot we can see that peaks in the  SLCC and the SLRCC do not align with each other, their separation defines the region of coexistence of both strains.  
} \label{fig:multilayer}
\end{center}
\end{figure}

As a numerical example consider the case where all edge topologies follow a Poisson distribution such that the number of $\tau$ edges is $\eta_\tau$ then
\begin{equation}
    p_\text{or}(\eta_\bot,\eta_{\bot,\text{og}}) = \frac{\langle \eta_\bot\rangle ^{\eta_\bot}e^{-\langle \eta_\bot\rangle}}{\eta_\bot!}\frac{\langle \eta_{\bot,\text{og}}\rangle ^{\eta_{\bot,\text{og}}}e^{-\langle \eta_{\bot,\text{og}}\rangle}}{\eta_{\bot,\text{og}}!}
\end{equation}
and 
\begin{equation}
    p_\text{gr}(\eta_\Delta,\eta_{\bot,\text{og}}) = \frac{\langle \eta_\Delta\rangle ^{\eta_\Delta}e^{-\langle \eta_\Delta\rangle}}{\eta_\Delta!}\frac{\langle \eta_{\bot,\text{go}}\rangle ^{\eta_{\bot,\text{go}}}e^{-\langle \eta_{\bot,\text{go}}\rangle}}{\eta_{\bot,\text{go}}!}
\end{equation}
The expected outbreak size of the first epidemic on the orange layer is then 
\begin{equation}
    S_\text{o}=1-e^{g_\bot(\langle\eta_\bot\rangle-1)}e^{g_{\bot,\text{og}}(\langle\eta_{\bot,\text{og}}\rangle-1)}
\end{equation}
while the green layer has 
\begin{equation}
    S_\text{g}=1-e^{g_\Delta(\langle\eta_\Delta\rangle-1)}e^{g_{\bot,\text{go}}(\langle\eta_{\bot,\text{go}}\rangle-1)}
\end{equation}
The $g_\tau$ equations for each are given by Eqs \ref{eq:gbot} and \ref{eq:gtriangle} for the intralayer tree-like and triangle edges, respectively. The interlayer tree-like connections have a subtle symmetry breaking depending on which layer we consider the focal node to belong to. We define 
\begin{equation}
    g_{\bot,\text{og}}(u_{\bot,\text{go}};T_1) = u_{\bot,\text{go}}+(1-u_{\bot,\text{go}})(1-T_1)
\end{equation}
and 
\begin{equation}
    g_{\bot,\text{g0}}(u_{\bot,\text{og}};T_1) = u_{\bot,\text{og}}+(1-u_{\bot,\text{og}})(1-T_1)
\end{equation}
since, each focal node depends on the other end being uninfected. Each $u_\tau$ is then the solution to a self-consistent equation according to Eq \ref{eq:G1_1}. 

The outbreak of the second epidemic follows from section \ref{subsec:S2} and in the Poisson case is \begin{equation}
    S_{2,\text{o}}=1-e^{h_\bot(\langle\eta_\bot\rangle-1)}e^{h_{\bot,\text{og}}(\langle\eta_{\bot,\text{og}}\rangle-1)}
\end{equation}
while the green layer has 
\begin{equation}
    S_{2,\text{g}}=1-e^{h_\Delta(\langle\eta_\Delta\rangle-1)}e^{h_{\bot,\text{go}}(\langle\eta_{\bot,\text{go}}\rangle-1)}
\end{equation}
We examine this system in Fig \ref{fig:multilayer}. The network is constructed such that the clustering coefficient of the green layer is $C=1/3$ with mean degree $\langle k_\Delta\rangle=6$ while the orange layer is $C=0$ with mean tree-like degree $\langle k_\bot\rangle=3.3$; a small number of interlayer edges were then added to connect the layers. In our experiment, the green-layer undergoes its phase transition at a lower $T_1$ than the orange layer due to its clustering. This causes the outbreak fraction of the first strain to show a double 2nd-order transition \cite{PhysRevX.4.041020,2020arXiv200606744M}. We confirm the presence of a phase transition by plotting the experimental second largest connected component (SLCC), peaks in which indicate a critical point.

Due to the different connectivity of each layer, the RG also experiences two critical points. We confirm this by plotting the second largest residual connected component (SLRCC), peaks in which indicate the presence of a phase transition in the residual network. The difference between the first peak in the SLCC and the last peak in the SLRCC defines the transmissibility range that allows coexistence of each strain in the network.

\section{Conclusion}
\label{sec:conclusion}

The study of disease spreading among human contact networks is of fundamental importance to society.
In particular, the study of multiple sequential strains with the presence of clustering can provide realistic models of social interactions capable of pathogen transmission.
In this paper, we have studied the problem of bond percolation on the RG of clustered configuration networks created by a prior bond percolation process. This represents two strains spreading sequentially among a population. 

We investigated the expected outbreak sizes of each epidemic as a function of the clustering coefficient of a substrate Poisson contact network with fixed average degrees. We found that networks with higher clustering coefficients exhibit reduced epidemic thresholds, $T_{1,c}$, of the first strain as well as smaller outbreak sizes; therefore, having a dual effect on $S_1$ parameterised by $T_1$. Networks with larger $\mathcal C$ values was found to reduce the maximum outbreak size of the second strain. The largest value of $T_1$ that permits the spreading of the second strain, $T_{1}^*$, is also reduced by clustering for these networks. This indicates that increased clustering forces the second strain to occupy a smaller region of the model's phase space and thus reduces its ability to become epidemic. The phase region that permits the coexistence of each strain, measured by the difference between $T_{1,c}$ and $T_1^*$, is also reduced with increased $\mathcal C$. Initially, this region broadens with the introduction of triangles to the contact network ($T_\delta$ in plot (B) of Fig \ref{fig:together}). However, the loss of tree-like edges causes the RG to fracture more than the original network when clustering is present, as shown by plotting $T_\delta$. 

We then applied the model to the uniform-degree model \cite{miller_2009} and found that clustering can be shown to increase the coexistence threshold of the second strain in addition to increasing the outbreak size as a function of $T$, relative to the configuration model. These results, in opposition to the findings of the Poisson experiment, suggest that it is the degree-assortativity of the residual graph that is the key factor in the progression of strain-2 to become an epidemic. In particular, the reduction of the outbreak size and the reduction of the coexistence threshold observed in the random Poisson experiments are due to the tendency of high $\mathcal C$ networks to assort their contacts by degree, and is not due to clustering.

We applied this model to the study of multilayer networks providing a numerical example of a 2-layer system. We found that the presence of a double 2nd-order phase transition in the GCC also creates a double 2nd-order phase transition in the GRCC. This was supported by examining the structure of the SLRCC as a function of the tranmissibility of the first strain.  

The results presented here suggest that the clustering of contacts can increase the epidemic threshold of the first disease and also reduce its outbreak size compared to independent edges.
However, under a perfectly cross-immune coupling, this has a negative impact for subsequent strains of the disease; enabling and aiding their proliferation.  
There is clearly an urgent need to study other possible interactions between strains in order to provide a theoretical framework within which to study the effects of different disease countermeasures, which may exhibit significantly different efficacies in different interaction regimes.

\subsection*{References}

\bibliography{bib}

\end{document}